\documentclass[twocolumn,showpacs,preprintnumbers,amsmath,amssymb]{revtex4}

\usepackage{graphicx}
\usepackage{bm}
\usepackage{color}

\definecolor{vitorob}{rgb}{0.1,0.45,0.80}

\begin{document}

\title{Flow graphs: interweaving dynamics and structure}


\author{R. Lambiotte$^{1}$\email{r.lambiotte@imperial.ac.uk}, R. Sinatra$^{2,3}$, J.-C. Delvenne$^{4,5}$, T.S. Evans$^{6}$,  M. Barahona$^{1}$ and V. Latora$^{2,3}$}

\affiliation{
$^1$ Institute for Mathematical Sciences, Imperial College London, 53 Prince's Gate,  SW7 2PG London, UK\\
$^2$ Dipartimento di Fisica e Astronomia, Universit\`a di Catania and INFN, Via S. Sofia 64, 95123 Catania, Italy\\
$^3$ Laboratorio sui Sistemi Complessi, Scuola Superiore di Catania, Via S. Nullo 5/i, 95123 Catania, Italy\\
$^4$ D{\' e}partement de Math{\' e}matique, Facult{\' e}s Universitaires Notre-Dame de la Paix, B-5000 Namur, Belgium\\
$^5$ Naxys, Facult{\' e}s Universitaires Notre-Dame de la Paix, B-5000 Namur, Belgium\\
$^6$ Theoretical Physics, Imperial College London, SW7 2AZ, U.K
} 

\begin{abstract} 
The behavior of complex systems is determined not only by the topological organization of their interconnections but also by the dynamical processes taking place among their constituents. A faithful modeling of the dynamics is essential because different dynamical processes may be affected very 
differently by network topology. A full characterization of such systems thus requires a formalization that encompasses both aspects simultaneously, rather than relying only on the topological adjacency matrix. To achieve this, we introduce the concept of flow graphs, namely weighted networks where dynamical flows are embedded into the link weights. Flow graphs provide an integrated representation of the structure and dynamics of the system, which can then be analyzed with standard tools from network theory. Conversely, a structural network feature of our choice can also be used as the basis for the construction of a flow graph that will then encompass a dynamics biased by such a feature.  We illustrate the ideas by focusing on the mathematical properties of generic linear processes on complex networks that can be represented as biased random walks and also explore their dual consensus dynamics. 
\end{abstract}


\maketitle

{\bf Introduction.}  The last decade has witnessed an explosion in the number of metrics for the characterization of complex networks \cite{review,bocca}. 
Most of these quantities rely on the analysis of topological properties and are, in a sense, combinatorial as they count certain motifs, e.g. edges, triangles, shortest paths, etc. Since this kind of measures do not account for patterns of flow on the network, 
flow-based metrics have also been proposed \cite{batty,rosvall,LDB08,delvenne,evans,rosvall2} and shown to provide radically new insight, especially in directed networks. However, these metrics usually have the limitation to be defined for discrete-time, unbiased random walks, which might not represent a good description for the process taking place on the graph under scrutiny. Among the systems where unbiased random walks are not realistic, let us mention the Internet and traffic networks, where a bias is necessary to account for local search strategies and navigation rules \cite{wang,fronczak,lee,tadic,zlatic}. Whenever complex inter-dependences between network sub-units are generated by patterns of flow \cite{rosvall2}, e.g. information in social networks or passengers in airline networks, neglecting or mis-interpreting the dynamics taking place on the graph leads to an incomplete and sometimes misleading characterization of the system. 

The main purpose of this work is to develop a mathematical framework that allows to analyze the structure of complex networks also from a dynamical point of view. To do so, we focus on a broad range of linear processes, namely biased random walks and consensus dynamics. We show how to define an alternative representation of the graph, called \emph{flow graph}, which naturally embeds flows in the weight of the links and on which dynamical processes become unbiased. In this way, to the same topological graph one can associate many different flow graphs, each specific of the different dynamics under consideration. This emphasizes the idea that the same original graph may exhibit different patterns of flow depending on the underlying dynamics, and that the choice of a metric as well as the extraction of pertinent information from a network should be made according to the nature of the dynamical process actually taking place on it.

In the following, we focus on undirected networks $\mathcal{G}$, which are described by their $N \times N$ symmetric adjacency matrix $A$, where $N$ is the number of nodes. By definition, $A_{ij}$ is the topological weight of the edge going from $j$ to $i$. The strength $s_i=\sum_j A_{ij}$ of node $i$ is the total weight of the links connected to it. If the network is unweighted, $s_i$ is simply the degree of node $i$.  $W=\sum_{ij} A_{ij}/2$ is the total weight in the network. Whereas the adjacency matrix reflects the underlying topology, nothing so far determines the dynamical processes operating on the system \cite{batty}. Here, we consider a broad class of linear processes defined by the equation:
\begin{equation}
\label{processes}
x_{i;t+1} = \sum_j B_{ij} x_{j;t}
\end{equation}
where the evolution of a quantity $x_i$, associated to node $i$, is driven by $B_{ij}$, a matrix related to the adjacency matrix $A_{ij}$. In particular, in the following we will focus on two subclasses of (\ref{processes}), namely \emph{random walks} and \emph{consensus problems}.

\smallskip

{\bf Flow graphs for general random walks.} We start our discussion with dynamical processes aiming at modeling the diffusion of some quantity or information on $\mathcal{G}$. The simplest process we can consider is a discrete-time, unbiased random walk (URW) where, at each step, a walker located at a node $j$ follows one of the links of $j$ with a probability proportional to its weight. In this case, the expected density of walkers at node $i$, denoted by $p_i$, evolves according to the rate equation
\begin{equation}
\label{discrete}
p_{i;t+1} = \sum_{j} T_{ij} p_{j;t},
\end{equation}
where $T$ is the transition matrix whose entry $T_{ij}$ represents the probability to jump from $j$ to $i$
\begin{equation}
\label{unbiased}
T_{ij}=A_{ij}/s_j \;.
\end{equation}
In order to preserve the total number of walkers, $T_{ij}$ satisfies the condition to be column normalized,  i.e. $\sum_{i} T_{ij}=1$. Consequently, $\sum_i p_{i;t}=1$ is verified for every $t$.
The dynamical process (\ref{discrete}) with transition matrix (\ref{unbiased}) is known to converge to the equilibrium solution 
$p^{*}_i=s_i/2W$
 if the graph is connected and non-bipartite, i.e. if the dynamics is ergodic \cite{chung0}.

{\it Biased random walks.} There exist infinitely many other ways to define a random walk and thus to model diffusion on the same graph $\mathcal{G}$. An interesting class of processes are biased random walks (BRWs), defined as follows \cite{latora}. Let each node $i$ be given a definite positive attribute $\alpha_i$. Then a walker located at node $j$ decides to jump onto one of its neighbors, say $i$, with a probability proportional to $\alpha_i A_{ij}$. Hence, the probability to jump from $j$ to $i$ is given by 
\begin{equation}
\label{biased}
 {T}^{(\alpha)}_{ij}= \frac{ \alpha_i A_{ij}}{\sum_k  \alpha_k A_{kj}}.   
\end{equation}
This is equivalent to saying that the motion of a walker is biased according to the values of  $\alpha$ associated to the nodes. The attribute $\alpha_i$ can be either a topological property of node $i$, such as its strength $s_i$ or its betweenness centrality, or, more in general, can represent an arbitrary function of an intrinsic node property, as for instance the reputation of a person in a social network. For different $\alpha$, BRWs correspond to distinct diffusive processes characterized by different spectral properties for (\ref{biased}).

Let us show that it is always possible to interpret the BRW defined by (\ref{biased}) as an URW on an opportunely defined flow graph $\mathcal{G}^{'}$ \footnote{This result holds for BRWs where a symmetric attribute $\alpha_{ij}=\alpha_{ji}$ is assigned to edges instead of to nodes.}. This observation has important implications, as it makes possible to use theoretical results known for URWs for the analysis of BRWs. In addition to this, as we will develop below, this representation supplies an alternative, advantageous way to highlight dynamical characteristics of the system.
Let us define the non-negative and symmetric matrix 
\begin{equation}
\label{x}
A^{'}_{ij} =\alpha_i A_{ij} \alpha_j.
\end{equation}
This is the adjacency matrix of the {\em flow graph} $\mathcal{G}^{'}$, whose edges are the same as in $\mathcal{G}$ but with different weights (see Fig.~1). It is straightforward to show that an URW on $\mathcal{G}^{'}$, described by the equation $p^{'}_{i;t+1} = \sum_{j} T^{'}_{ij} p^{'}_{j;t}$ with $T^{'}_{ij}=A^{'}_{ij}/s^{'}_j$, coincides with a BRW on $\mathcal{G}$ driven by the transition matrix (\ref{biased}), since ${T}^{(\alpha)}_{ij}\equiv T^{'}_{ij}$. Thus the equilibrium solution of the BRW on the original graph is given by
\begin{equation}
\label{brwstat}
p_j^{*'}=\frac{s^{'}_j}{2W^{'}} = \frac{\sum_i \alpha_i A_{ij} \alpha_j}{\sum_{i,j} \alpha_i A_{ij} \alpha_j},
\end{equation}
 in agreement with \cite{latora}. This result also shows that $A^{'}_{ij}$ is proportional to the flow of probability from $j$ to $i$ at equilibrium \cite{LDB08}. 
 
 \begin{figure}
\includegraphics[width=0.4\textwidth]{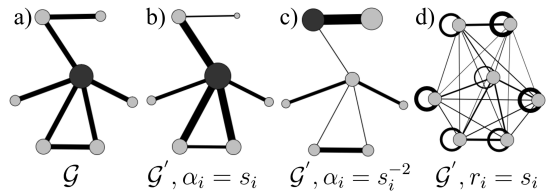}
\vspace{-0.3cm}
\caption{Visual representation of an unweighted graph $\mathcal{G}$ (a) and of its flow graphs $\mathcal{G}^{'}$ defined for BRWs with attributes $\alpha_i=s_i$ (b) and $\alpha_i=s_i^{-2}$ (c), and a continuous-time random walk with $t=2, r_i = s_i$ (d). The width of the links is proportional to their weight, and the surface of the nodes to their strength. The strength leader, i.e. the node with the highest strength, is darkened if it exists. 
In order to make the graphs comparable, we renormalize the weights in order to ensure that $W=W^{'}$. These examples clearly show that different dynamics lead to different patterns and that important nodes for one dynamics might be less important for other dynamics.}\label{fig1}
\end{figure}

In order to illustrate these concepts, let us focus on a class of BRWs where  $\alpha_i$ has a power-law dependence on the strength, $\alpha_i = s_i^\gamma$. This functional dependence has been proposed by several authors in order to model local routing strategies \cite{wang,fronczak,latora}. By changing the exponent $\gamma$, one tunes the dependence of the bias on the strength. When $\gamma=0$, the standard URW is recovered, while biases toward high (low) strengths are introduced when $\gamma>0$ ($\gamma<0$). From (\ref{brwstat}), one finds
\begin{equation}
\label{onebias}
p^{*'}_j=\frac{\sum_i s_i^\gamma A_{ij} s_j^\gamma}{\sum_{i,j} s_i^\gamma A_{ij} s_j^\gamma},
\end{equation}
which emphasizes that the equilibrium density of walkers at $j$ now depends on the strength of $j$ and of its neighbors for any  $\gamma \neq 0$. In the heterogeneous mean-field approximation where the adjacency matrix is factorized $A_{ij} \approx s_i s_j/2W$, one recovers  the known expression 
$p^{*'}_j=s_j^{\gamma+1}/(N \langle s_j^{\gamma+1}\rangle)$ \cite{LDB08,wang,fronczak}.
 
Another interesting class of BRWs is one where bias is performed towards high eigenvector centrality node \cite{parry,demetrius,delvenne0,burda,latora,optimal}, $\alpha_i=v_i$, where $v$ is the dominant eigenvector of $A$  \cite{bonacich}, namely $\sum_j A_{ij} v_j= \lambda_1 v_i$ and $\lambda_1$ is the largest eigenvalue. This bias leads to the maximal-entropy random walk defined by
\begin{equation}
\label{optimal}
p_{i;t+1} = \sum_{j} \frac{v_i A_{ij}}{\lambda_1 v_j} p_{j;t},
\end{equation}
which is known to be maximally dispersing on the graph \cite{parry}, in the sense that the entropy rate is optimal. By defining a flow graph whose adjacency matrix has the form $A^{'}_{ij} = v_i A_{ij} v_j$, an URW on $A^{'}$ exhibits a stationary probability distribution which is also the solution of (\ref{optimal}), i.e. $p_i^*=v_i^2/Z$, with $Z=\sum_i v_i^2$.

{\it Continuous-time random walks.} When modeling diffusion, a broad range of processes opens up if walkers can perform their jumps asynchronously. A natural way to implement this situation is to switch from a discrete-time to a continuous-time perspective \cite{montroll}, which finds many applications in biological and physical systems. Passage to continuity can be done in many ways, each leading to a different stochastic process. In the following, we restrict the scope to Markovian processes where the waiting times between two jumps are Poisson distributed. Without loss of generality, we also assume that walkers jump in an unbiased way, while keeping in mind that any BRW can be seen as an unbiased process on the associated flow graph. The time-interval between two jumps is determined by the so-called waiting time distribution $\psi(i;t)=r_i e^{- r_i t}$. The rate $r_i$ at which walkers jump may in general be non-identical and depends on the node $i$ where the walker is located. Different sets of $\left\lbrace r_i \right\rbrace$ generate different stochastic processes, though the sequence of nodes visited $i_{0},i_{1},....,i_{\tau}$, where $i_{\tau}$ is the node visited after $\tau$ jumps,does not depend on the $\left\lbrace r_i \right\rbrace$. For different choices of $\left\lbrace r_i \right\rbrace$, what changes is only the times at which the jumps are performed and the time intervals spent on the nodes.

Such continuous-time random walks are driven by the rate equation
 \begin{equation}
\label{ctrwgeneral}
\dot{p}_{i} = \sum_{j} \left( \frac{A_{ij}}{s_j} r_j  - r_i \delta_{ij}\right) p_j \equiv - \sum_{j} L_{ij} p_j
\end{equation}
whose stationary solution $p_i^{*}=s_i/(Z r_i)$, with $Z=\sum_i s_i/r_i$, can be intuitively understood as the probability to arrive at a node times the characteristic time $\approx 1/r_i$ spent on it. Standard choices for the jumping rates include the uniform rate $r_i=1$ $\forall i$, and the strength-proportional rate $r_i=s_i$ $\forall i$, for which one recovers the standard forms of the Laplacian operator $L_{ij}=\delta_{ij}-A_{ij}/s_j$ and $L_{ij}= s_i \delta_{ij} - A_{ij}$ respectively \cite{LDB08}. 

This continuous-time random walk can also be viewed as a discrete-time URW, i.e. 
$p^{'}_{i;t} = \sum_{j} T^{'}_{ij} p^{'}_{j;0}$,
on a flow graph defined by the adjacency matrix 
\begin{equation}
\label{xx}
A^{'}_{ij}(t)=\left(e^{-t L} \right)_{ij}  Zp_j^{*} = \left(e^{-t L} \right)_{ij} \frac{s_j}{r_j}.
\end{equation}
The definition (\ref{xx}) follows from the solution of equation (\ref{ctrwgeneral}) which gives $p_i(t)=\sum_j \left( e^{-tL} \right)_{ij} p_j(0)$. In fact, one can interpret the probability distribution of a continuous time-random walk at time $t$ as the result of one step random walk driven by the transition matrix $T^{'}_{ij}(t)=A'_{ij}(t)/s^{'}_j$. 
As previously, $A^{'}_{ij}(t)$ is the flow of probability from $j$ to $i$ at stationarity.
One easily verifies that $A^{'}_{ij}(t)$ is symmetric due to detailed balance, i.e. $\sum_j T^{'}_{ij} p^{'*}_j=\sum_i T_{ij} p^{'*}_i$ at equilibrium, and that $\sum_j A^{'}_{ij}(t)=\frac{s_i}{r_i}$ at all times. The associated flow graph naturally summarizes how random walkers probe the network over a certain time scale and provides a representation of the system over this scale \cite{LDB08}.

\smallskip
{\bf Consensus processes.}  Another kind of interesting processes belonging to the class (\ref{processes}) is the so-called ``distributed consensus'', for which nodes imitate their neighbors such as to reach a uniform, coordinated behavior.
In its simplest form, consensus dynamics is implemented by the so-called agreement algorithm \cite{tsitsi}. Each node $i$ is endowed with a scalar value $x_i$ which evolves as 
\begin{equation}
\label{inverse}
x_{i;t+1}=\frac{1}{s_i} \sum_j A_{ij} x_{j;t}.
\end{equation}
At each time step, the value on a node is updated by computing a weighted average of the values on its neighbors. If the graph is connected and non-bipartite, consensus is asymptotically achieved and each node reaches the uniform value $x^{*}=\sum_i  x_{i;0} s_i/(2W)$ given by a weighted average of the initial conditions. The agreement algorithm (\ref{inverse}) is different from an URW, e.g. it does not conserve $\sum_i x_i$ except if the graph is regular. Nonetheless, it has the interesting property to be dual of the URW, as it is driven by the transpose of (\ref{unbiased}) \cite{batty,krause}. Moreover, both processes can be seen as two interchangeable facets of the same dynamics, as their spectral properties are related by a trivial transformation, namely left and right eigenvectors of (\ref{unbiased}) are related by $v^{R}_{\alpha;i} = s_i v^{L}_{\alpha;i}$, where $\alpha \in[1,N]$ is an index over the eigenvectors. 

Similarly to the URW, (\ref{inverse}) can be generalized either by introducing a bias in the weighted average or by tuning the rate at which nodes compute the average of their neighbors' values. The broad class of consensus dynamics generated by this scheme includes for instance models from opinion dynamics \cite{krause} and linearized approach to synchronization of different variants of the Kuramoto model \cite{pecora,motee,arenas,motter}. However, what is most important is that, for any bias, the duality to the random walk (\ref{discrete}) allows to introduce a consensus as (\ref{inverse}) on the associated flow graph, analogous to (\ref{x}).

\smallskip

{\bf Discussion.} 
The behavior of complex systems is determined by their structure and their dynamics \cite{batty}. A purely structural analysis, where properties of the adjacency matrix are considered without any insight on underlying dynamical processes, provides only a partial understanding of the system. 
In this paper, we have focused on a broad range of linear processes on networks. Some examples where this kind of processes are used is for modeling diffusion or synchronization, and they all exhibit distinct dynamical properties. These properties are summarized by their associated flow graph $\mathcal{G}^{'}$, where the weight of a link is dictated by the patterns of dynamical flow at equilibrium. The definition of $\mathcal{G}^{'}$ has the advantage of simultaneously representing the network topology and its dynamics, and of properly emphasizing nodes and edges which are important from a dynamical point of view. 
As shown in Fig.~1, details of the underlying dynamics strongly affect the importance of nodes and their associated ranking \cite{delvenne0}. Standard network metrics can be measured on the flow graph in order to uncover other aspects of its dynamical organization \cite{ott}, for instance to measure centrality for BRWs \cite{lee}.

An important context where our formalism proves useful is community detection \cite{zlatic}. The modular structure of a network is often uncovered by optimizing a quality function for the partition $\mathcal{P}$ of the nodes into communities \cite{F09}. The widely-used modularity \cite{newman} measures if links are more abundant within communities than would be expected on the basis of chance. Because of its combinatorial nature, modularity is known to be insensitive to important structural properties which may constraint a flow taking place on the network \cite{rosvall}. Alternative quality functions have thus been developed based on the idea that a flow of probability should be trapped for long times in communities when the partition is good \cite{rosvall,LDB08,delvenne}. An interesting quantity is the so-called stability $R(t)$  \cite{delvenne} which is defined as the probability for a random walker to be in the same community initially and at time $t$, when the system is at stationarity. Stability is in general different from modularity, but they coincide when the random walk is discrete-time and unbiased, the network undirected and $t=1$. The notion of flow graph naturally reconciles combinatorial and flow-based approaches, as the stability of a graph for any process is equal to the modularity of its corresponding flow graph \cite{LDB08}, and allows for the detection of modules adapted to the system under scrutiny.
In systems where dynamical processes are known to differ from URWs, the notion of flow graph thus provides the means to apply standard combinatorial methods while still properly taking into account the dynamical importance of nodes and links. 

The equivalence between trajectories of a biased (or continuous-time) random walker on $\mathcal{G}$ and those of an URW on $\mathcal{G}^{'}$ also has important practical implications, as it allows to make use of well-known theoretical results to analyze BRW processes, for instance their stationary solution and conditions to convergence, mean first-passage time \cite{rwc,benichou} or spectral properties \cite{chung}. This theoretical framework might prove useful to address several problems related to BRWs, such as the search of local biases $\alpha_i$ optimizing in some way the performance of the system \cite{optimal}, for instance by balancing load on the nodes and improving search in routing systems \cite{randles,lee}, or enhancing the prediction of missing links in empirical data-sets \cite{jure}.

{\bf Acknowledgements}
R.L. has been supported by UK EPSRC. This work was conducted under the HPC-EUROPA2 project (project number: 228398) with the support of the European Commission Capacities Area - Research Infrastructures Initiative.

\end{document}